\documentclass[lettersize,journal]{IEEEtran}
\usepackage{amsmath,amsfonts}
\usepackage{amsthm}
\usepackage{algorithmic}
\usepackage{algorithm}
\usepackage{array}
\usepackage[caption=false,font=normalsize,labelfont=sf,textfont=sf]{subfig}
\usepackage{textcomp}
\usepackage{stfloats}
\usepackage{url}
\usepackage{verbatim}
\usepackage{graphicx}
\usepackage{cite}
\usepackage{amssymb}
\newtheorem{theorem}{Theorem}
\newtheorem{lemma}{Lemma}

\newtheorem{corollary}{Corollary}
\newcommand{\R}{{\mathbb{R}}}

\begin{document}

\title{A Multiple Artificial Potential Functions Approach for Collision Avoidance in UAV Systems}

\author{Oscar F. Archila$^{1,2}$, Alain Vande Wouwer$^{2}$ and  Johannes Schiffer$^{3}$ 
\thanks{*This work is partially supported by the BMBF project iCampus2 (\#16ME0420K).}
\thanks{$^{1}$Oscar F. Archila is with the Control Systems and Network Control Technology Group, Brandenburg University of Technology Cottbus-Senftenberg (BTU C-S), 03046 Cottbus, Germany and with the Department of System, Estimation, Control and Optimization (SECO), University of Mons, 7000 Mons, Belgium
        {\tt\small archilac@b-tu.de}}%
\thanks{$^{2}$Alain Vande Wouwer is with the Department of System, Estimation, Control and Optimization (SECO), University of Mons, 7000 Mons, Belgium
	{\tt\small alain.vandewouwer@umons.ac.be}}
\thanks{$^{3}$J. Schiffer is with the Control Systems and Network Control Technology Group, Brandenburg University of Technology Cottbus-Senftenberg (BTU C-S), 03046 Cottbus, Germany and the Fraunhofer IEG, Fraunhofer Research Institution for Energy Infrastructures and Geothermal Systems IEG, 03046 Cottbus, Germany {\tt\small schiffer@b-tu.de}}
}
\markboth{Journal of \LaTeX\ Class Files,~Vol.~14, No.~8, August~2021}%
{Shell \MakeLowercase{\textit{et al.}}: A Sample Article Using IEEEtran.cls for IEEE Journals}

\IEEEpubid{0000--0000/00\$00.00~\copyright~2021 IEEE}

\maketitle

\begin{abstract}
Collision avoidance is a problem largely studied in robotics, particularly in unmanned aerial vehicle (UAV) applications. Among the main challenges in this area are hardware limitations, the need for rapid response, and the uncertainty associated with obstacle detection. Artificial potential functions (APOFs) are a prominent method to address these challenges. However, existing solutions lack assurances regarding closed-loop stability and may result in chattering effects. Motivated by this, we propose a control method for static obstacle avoidance based on multiple artificial potential functions (MAPOFs). We derive tuning conditions on the control parameters that ensure the stability of the final position. The stability proof is established by analyzing the closed-loop system using tools from hybrid systems theory. Furthermore, we validate the performance of the MAPOF control through simulations, showcasing its effectiveness in avoiding static obstacles. 
\end{abstract}


\section{Introduction}
\IEEEPARstart{T}{}he use of robotics in industrial, construction, and manufacturing processes has increased significantly with the introduction of new technologies that allow complex tasks to be performed more efficiently and autonomously, both in a stationary and mobile setting. For mobile applications, robotics systems formed by unmanned aerial vehicles (UAVs) have gained popularity in recent years \cite{molina2023review}. 
They are used for various applications, such as disaster management, remote sensing, search and rescue, and precision agriculture \cite{mohsan2023unmanned}. However, reliably integrating UAVs into the aforementioned fields of application poses several new challenges, such as data security, communication uncertainties, restricted power consumption, complete autonomy, and safe operation \cite{messaoudi2023survey}. The safe operation of UAVs is a high-priority concern {\cite{wackwitz2015safety}.

 To address this safety concern, in mobile robotics applications, typically, the robots' local control is enhanced by collision avoidance systems (CAS) \cite{yasin2020unmanned}. A CAS is a dedicated control algorithm with great reaction speed and adaptation to variable environments to increase the safe operation of robotic systems in uncertain environments. Given its relevance, the design of CAS is a highly active area of research  \cite{yasin2020unmanned,huang2019collision}. Some of the main open challenges of CAS design are deadlock (DL), motion planning in uncertain environments, and the lack of theoretical guarantees in the closed-loop CAS performance \cite{huang2019collision}. 
 
Within the broad field of CAS schemes, two main directions can be identified: barrier function-based approaches and strategies based on artificial potential functions (APOFs) \cite{singletary2021comparative,yasin2020unmanned,huang2019collision}. APOFs are one of the most studied methods to deal with the CAS problem in robotics due to their simple implementation and low computational cost for real-time applications \cite{Ge2000, Urakubo2004}. As the name suggests, the design of APOFs is inspired by applying additional "forces" to the low-level tracking control of the UAV to avoid obstacles while following a trajectory or reaching a final target position. 
For instance, in \cite{garcia2012quad}, the authors integrate APOFs into UAVs to avoid fixed obstacles while maintaining a desired formation. APOFs are also applied in various other environments, such as dynamic 3D environments \cite{Chen2013} or indoor environments~\cite{Mac2016}. 

Despite its popularity, CAS based on APOFs suffers from two main drawbacks: chattering and the lack of tuning criteria that ensure closed-loop stability. The chattering problem has been addressed by implementing improvements in APOFs, where the direction in which the obstacle is evaded is more appropriately selected \cite{sun2017collision,weerakoon2015artificial}. Furthermore, geometric concepts are integrated to generate a smoother trajectory while avoiding obstacles in 3D \cite{Zhang2018}. However, none of the previously mentioned APOFs approaches guarantee closed-loop stability. Furthermore, this strategy suffers from several other drawbacks, such as oscillations in the modified reference trajectory and deadlocks~\cite{weerakoon2015artificial}.

These essential drawbacks can be overcome by the use of barrier functions. In \cite{Wang2016}, a barrier function-based methodology for multi-objective mobile robot teams is proposed to avoid obstacles while maintaining inter-agent communication. This approach has been extended to UAVs \cite{Xu2018, Restrepo2019, Zhang2020, Hegde2021}. 
Yet, the implementation and deployment of barrier function-based CASs is rather complex. Another main challenge with these algorithms is their high computational cost, which can be particularly limiting on small UAV platforms with restricted hardware, where real-time action is critical for successful obstacle avoidance \cite{bharati2018real}.
\newpage

Given these open challenges, we propose a Multiple APOF (MAPOF) scheme to avoid multiple stationary obstacles in UAV systems. This approach provides the simplicity and low computational cost of APOFs while ensuring rigorous stability guarantees. The latter is established using tools from switched systems theory. In summary, the three main contributions of this work are as follows.
	\begin{itemize}
		\item Inspired by the APOFs approach, we propose a CAS strategy based on Multiple APOFs (MAPOFs), i.e., several forces that switch depending on the respective positions of the UAV, obstacle, and target. As with APOFs, MAPOFs require low computational cost and exhibit a fast response.
		Compared to standard APOFs schemes, the proposed MAPOF strategy alleviates the appearance of unwanted phenomena, such as oscillations, chattering, and deadlocks. Formally, the proposed MAPOF scheme is a state- and time-dependent switched control law.
		\item We derive tuning conditions on the MAPOF parameters, which ensure closed-loop stability. This is achieved by combining the MAPOF control with second-order UAV dynamics and interpreting the resulting closed-loop system as a switched piecewise affine linear system. We then use the concept of a dwell-time \cite{morse1996supervisory} along with stability analysis tools from hybrid system theory \cite{decarlo2000perspectives} to derive rigorous conditions for stability.
		\item We provide extensive simulation results, which demonstrate how the application of MAPOFs allows the UAVs to avoid obstacles while assuring the stability of the closed-loop system.
	\end{itemize}

The remainder of this paper is organized as follows. In Section II, the problem is stated, and in Section III, the MAPOF control strategy is introduced. In Section IV, the tuning conditions for closed-loop stability are presented. The numerical results for two different scenarios are developed in Section V. Finally, conclusions and an outlook on future work are drawn in Section VI.\\

\section{PROBLEM FORMULATION}
\label{sec:prob}
The main objective of this study is the safe, automatic navigation of a UAV in a workspace while aiming to reach a specific target position. However, unexpected fixed obstacles may be found along the desired trajectory. Hence, it is necessary to design a CAS that ensures that these obstacles are avoided and that the UAV reaches its final position safely and in a timely manner. To this end, it is assumed that the UAV has obstacle detection capabilities and receives companion computing information, such as its current position and velocities. 

The primary objective of the UAV is defined as reaching the target position $\eta \in \mathbb{R}^2$. The current position of the UAV is represented as $\xi(t) \in \mathbb{R}^2$.  Following common practice, it is assumed that the UAV is equipped with a low-level control \cite{han2018review}. Hence, a holonomic second-order system ca be used for the UAV dynamics \cite{ren2008distributed}, i.e.,

\begin{equation}
	\ddot{\xi}=u.
	\label{eq:system_initial}
\end{equation}

The control variable $u(t) \in \mathbb{R}^2$ represents a high-level control for the desired trajectory of the UAV. The following assumptions are made:
\begin{itemize}
    \item The UAV detects an obstacle at a position $\zeta \in \Omega$ as soon as the Euclidean distance between the UAV and the obstacle is less than the detection radius $r_d \in \mathbb{R}_+$.
    \item The minimum distance between any two obstacles is $r_d$.  Consequently, the UAV can only be confronted with one obstacle at any point in time.
    \item The minimum distance between the target position and any obstacle is at least $r_d$.
    \item The UAV has continuous access to its position information, while information regarding the presence of obstacles is only available when an obstacle is within the detection radius $r_d.$
\end{itemize}
 The last consideration is that the UAV needs to calculate the control decision in real time to avoid obstacles effectively.

\section{MAPOF CONTROL STRATEGY}
A control strategy based on MAPOFs is designed to achieve the objectives formulated in Section~\ref{sec:prob}. The standard APOFs strategy suffers from severe drawbacks, such as oscillating trajectories and deadlocks \cite{weerakoon2015artificial}. To alleviate these issues, several forces are considered that switch depending on the respective positions of the UAV, obstacle, and target. This results in a state-dependent switched control law for the CAS.

\subsection{State space partitioning}
\label{subsec:partitioning}
 A partition of the state space $\mathbb{R}^2$ into four subregions is introduced, wherein a different artificial potential force is applied to the UAV. These subregions are calculated in real time when the UAV detects an obstacle. Let the current obstacle position be denoted by $\zeta\in\Omega$ and let $r_m >0$ with $r_d>r_m$ denote the radius of a circular security region around $\zeta$. Figure~\ref{fig:geo_diagram} illustrates the problem geometry and the following quantities: 

 \begin{figure}
     \centering
     \includegraphics[scale=0.6]{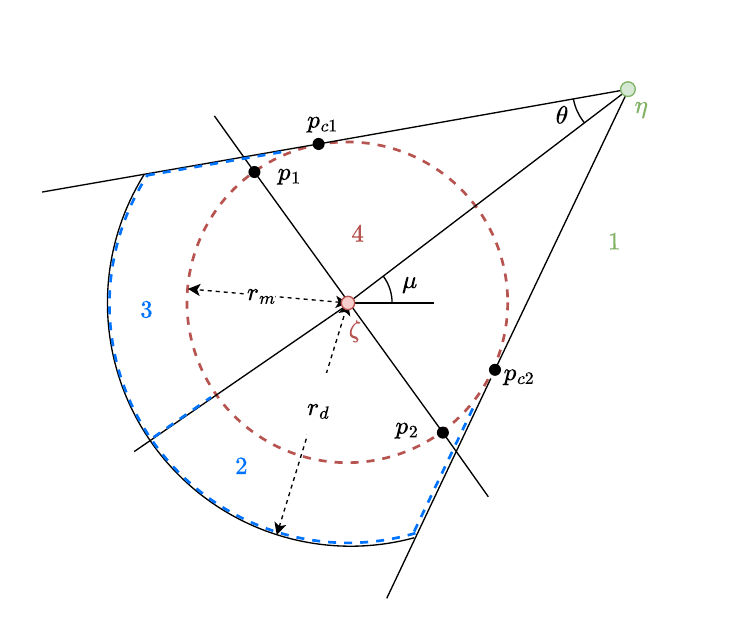}
     \caption{Introduction of geometric variables for the state space partitioning proposed in Section~\ref{subsec:partitioning}. The numbers indicate the respective subregion.}
     \label{fig:geo_diagram}
 \end{figure}

\begin{equation}
	\mu = \arctan \left (\frac{\eta_y-\zeta_y}{\eta_x-\zeta_x}  \right),
\end{equation}
\begin{equation}
	\theta = \arctan \left (\frac{r_m}{\left \| \eta-\zeta \right \|_2 }  \right),
\end{equation}
where $\mu \in \left [  0,2\pi \right)$ is the relative angle between the obstacle and the target, $\theta \in \left [  0,2\pi \right)$ is the relative angle between the target and the line that is projected from the target to the edge of the security region, $\left \| \cdot \right \|_2$ is the Euclidean norm, and the sub-indices $x$ and $y$ represent the first and second component of the vectors $\eta$ and $\zeta$.  

We define the points that intersect with the security zone perpendicularly to the obstacle position $\zeta$ as 

\begin{equation} 
	p_{1} = \left [  r_m  , 0 \right ] \textbf{R}\left(\mu+\frac{\pi}{2}\right) +\zeta,
\end{equation}
\begin{equation}
	p_{2} = \left [  r_m  , 0 \right ] \textbf{R}\left(\mu-\frac{\pi}{2}\right) +\zeta,
\end{equation}
where $\textbf{R}(\cdot)$ is the 2-D rotation matrix
\begin{equation*}
   \mathbf{R}(\phi)=\begin{bmatrix}
 \cos\phi&-\sin\phi \\ 
 \sin\phi& \cos\phi
\end{bmatrix}.
\end{equation*}
 
Similarly, the intersection points $p_{c1}$, $p_{c2} \in \mathbb{R}^2$ of the lines originating from the target position $\eta$ and the circle with radius $r_m$ and center $\zeta$ are defined as
\begin{equation}
	p_{c1} = \left [  r_m \cos(\theta) , r_m \sin(\theta) \right ] \textbf{R}(-\mu) +\zeta,
	\label{eq:pc1}
\end{equation}
\begin{equation}
	p_{c2} = \left [  r_m \cos(\theta) , -r_m \sin(\theta) \right ] \textbf{R}(-\mu) +\zeta.
	\label{eq:pc2}
\end{equation}
The state space $\Omega$ can be partitioned into four subregions as depicted in Figure~\ref{fig:geo_diagram}.

\subsection{Design of MAPOFs}
The proposed state space partitioning, see Figure~\ref{fig:geo_diagram}, suggests using four APOFs, instead of a single one. Our strategy is based on three criteria:
\begin{enumerate}
	\item When no obstacle is detected, an attraction force $F_1$ is induced to reach the target position.
	\item When an obstacle is detected, an avoidance force $F_2$ or $F_3$ is applied to change the UAV's direction.
	\item A repulsion force $F_4$ is applied when the UAV is close to the obstacle.
\end{enumerate}
Consequently, we introduce the following four forces
\begin{equation}
	\begin{split}
		&F_1(\xi)=k_{\eta}(\eta-\xi),\\	
		&F_2(\xi)=k_g(g_1^*-\xi),\\
		&F_3(\xi)=k_g(g_2^*-\xi),\\
		&F_4(\xi)=-k_\zeta(\zeta-\xi),
	\end{split}	
	\label{eq:forces}
\end{equation}
where $k_{\eta},k_g,k_{\zeta}>0$ and  $g_1^*\in\Omega$ as well as $g_2^*\in\Omega$ are virtual positions. The virtual positions $g_1^*$ and $g_2^*$ satisfy two constraints. They belong to the line generated by the points $p_{1}$ and $p_2$ and need to be outside regions 2 and 3. 

In Figure \ref{fig:map_force} the mapping of the forces $F_i,$ $i\in\{1,2,3,4\},$ given an obstacle $\zeta$ and a goal $\eta$ is shown. The numbers indicate the force used when the UAV is in the corresponding state space partition. 

\subsection{State- and time-dependent switching function}
The last step to complete the MAPOFs control strategy is the derivation of a switching function, which activates the correct APOF \eqref{eq:forces} in dependency of the current UAV position $\xi$, the obstacle position $\zeta$ and the target position $\eta$. To this end, let $z_c, \,z_1,\,z_2,\,\,z_3 \in \mathbb{R}^2$ and consider the function $\Gamma: \mathbb R^2\times \mathbb R^2\times \mathbb R^2\times \mathbb R^2\to \mathbb R,$
\begin{equation}
	\Gamma (z_c,z_1,z_2,z_3)= \begin{cases}
		1, & z_c \text{ is inside of the lines generated by} \\
		& z_1, z_2 \text{ and the origin } z_3 \\
		0, & \text{otherwise}.
	\end{cases}
\end{equation}

By using $\Gamma,$ we can now define the state switching function $\sigma_s:[0,\infty)\to\{1,2,3,4\}$ that activates the MAPOFs given in \eqref{eq:forces} as follows

\begin{equation}
	\sigma_{s} (t)=\begin{cases}
		4, &   \left \| \xi-\zeta \right \|_2 < r_m, \\
		3, & \Gamma(\xi,p_{c1},\zeta,\eta ) = 1,  \\
		& \Gamma(\xi-\zeta,(\eta-\zeta)\textbf{R}(\pi),p_{c1}-\zeta,[0,0]^\top)  =1, \\
		&  \left \| \eta-\zeta \right \|_2 > r_m , \, \left \| \eta-\zeta \right \|_2 < r_d, \\
		2,&\Gamma(\xi,p_{c2},\zeta,\eta ) = 1,  \\
		& \Gamma(\xi-\zeta,(\eta-\zeta)\textbf{R}(\pi),p_{c2}-\zeta,[0,0]^\top) = 1, \\
		& \left \| \eta-\zeta \right \|_2 > r_m, \,\, \left \| \eta-\zeta \right \|_2 < r_d, \\
		1, & \text{otherwise}.
	\end{cases}
	\label{eq:switching_signal_state}
\end{equation}

The function's return value indicates which of the four APOF in \eqref{eq:forces} is to be used: 

\begin{itemize}
    \item It returns 1, when no obstacle is detected.
    \item It returns 2 or 3, when the obstacle enters the detection area of the UAV given by $r_d$ and the UAV is going in the direction to collide.
    \item It returns 4, when the UAV is within the security region defined by $r_m$.
\end{itemize}

  \begin{figure}[t]
	\centering
	\includegraphics[scale=0.37]{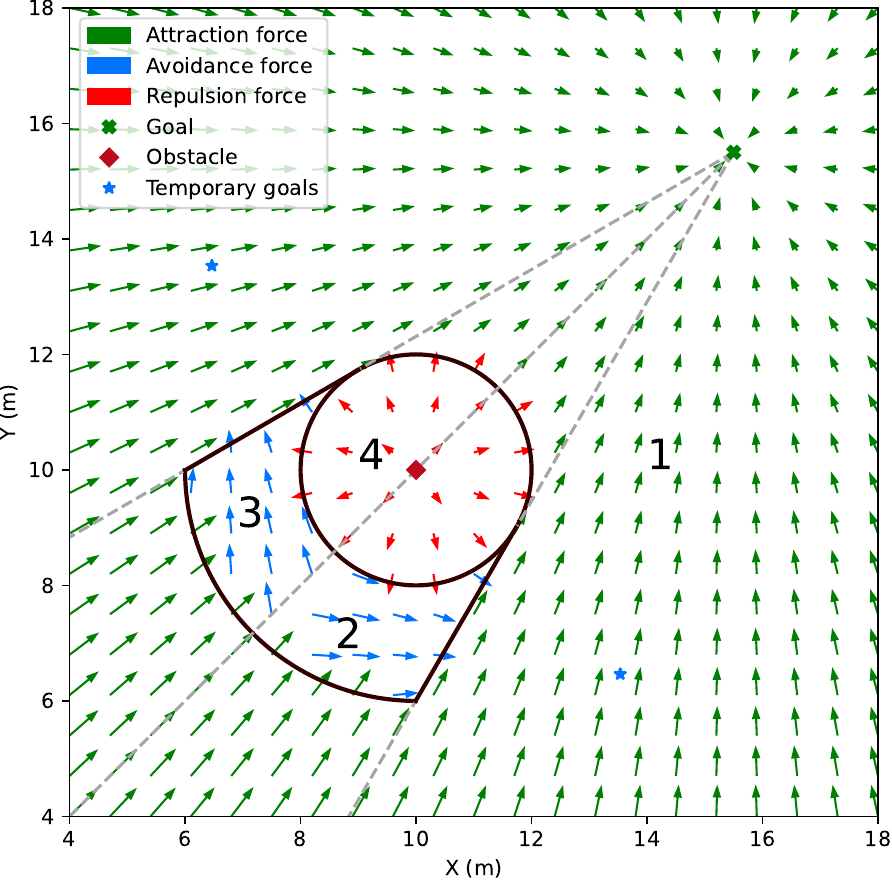}
	 \caption{Schematic representation of the MAPOF force map, with the four forces introduced in \eqref{eq:forces}, relative to the obstacle and goal positions.}
	\label{fig:map_force}
\end{figure}

To avoid chattering, we introduce a dwell-time \cite{morse1996supervisory}, which limits the number of switches within any time interval $T$. We define an infinite and strictly increasing sequence of switching times $\mathcal{T} =\{t_1,t_2,...,t_k,... \} $, with $k \in \mathbb{Z}_{>0}$. Consider the time interval $t_k >t \geq t_{k+1}$, where $t_k$ is the last switch, $t_{k+1}$ is the future switch, and the expression $t^-$ and $t_k^-$ the times just before $t$ and $t_k$. The state switching function in \eqref{eq:switching_signal_state} is complemented by adding a time-dependent switching restriction between subsystems to prevent chattering and generate a smooth path. The times $T_{D_1},T_{D_2} \in \mathbb{R}_{>0}$ restrict the state switching function \eqref{eq:switching_signal_state} as follows:
\begin{itemize}
    \item The switching time interval from any subsystem $\sigma\in\{1,2,3\}$ to any subsystem $\sigma\in\{1,2,3,4\}$ has to be greater than $T_{D_1}$.
    \item The switching time interval from subsystem $\sigma=4$ to any subsystem $\sigma\in\{1,2,3\}$ is not restricted, i.e., the switching can occur arbitrarily fast. 
    \item After a switch from subsystem $4$ to any subsystem $\sigma\in\{1,2,3\}$, the subsequent switch to any other subsystem $\sigma\in\{1,2,3\}$ has to be greater than $T_{D_2}>T_{D_1}$.
\end{itemize}

The state- and time-dependent rules lead to the switching function

\begin{equation}
	\sigma \left(\sigma_s,t\right) =\begin{cases}
	   4, &   \sigma_{s}(t)=4, \\
	   3, & \sigma_{s}(t)=3, \,\sigma_{s}(t^-)=4  \\
        & \textup{or} \, \bigr[\sigma_{s}(t_k)=\{1,2\}, \,  t-t_k>T_{D_1} \bigr]  \\
        & \textup{or} \, \bigr[\sigma_{s}(t_k^-)=4,\, \sigma_{s}(t_k)=\{1,2\},
        \\&\, t-t_k>T_{D_2}\bigr],\\
		2, & \sigma_{s}(t)=2, \,\sigma_{s}(t^-)=4  \\
        & \textup{or} \, \bigr[\sigma_{s}(t_k)=\{1,3\}, \,  t-t_k>T_{D_1} \bigr]  \\
        & \textup{or} \, \bigr[\sigma_{s}(t_k^-)=4,\, \sigma_{s}(t_k)=\{1,3\},
        \\&\, t-t_k>T_{D_2}\bigr],\\
		1,  & \sigma_{s}(t)=1, \,\sigma_{s}(t^-)=4  \\
        & \textup{or} \, \bigr[\sigma_{s}(t_k)=\{2,3\}, \,  t-t_k>T_{D_1} \bigr]  \\
        & \textup{or} \, \bigr[\sigma_{s}(t_k^-)=4,\, \sigma_{s}(t_k)=\{2,3\},
        \\&\, t-t_k>T_{D_2}\bigr],\\
        \sigma_{s}(t_k), & \text{otherwise}.
	\end{cases}
	\label{eq:switching_signal}
\end{equation}

A summary of the different cases for the MAPOFs activation function is presented in Table \ref{tab:switching_function}. 

\begin{table}[h]
\caption{Summary of the MAPOFs activation function $\sigma (t)$ in \eqref{eq:switching_signal}.}
\label{tab:switching_function}
\begin{tabular}{|c|c|c|c|c|c|}
\hline
\begin{tabular}[c]{@{}c@{}}Current \\ APOF $\sigma_s(t)$\end{tabular} & \begin{tabular}[c]{@{}c@{}}APOF \\ $\sigma_s(t^-)$\end{tabular} & \begin{tabular}[c]{@{}c@{}}APOF \\ $\sigma_s(t_k)$\end{tabular} & \begin{tabular}[c]{@{}c@{}}APOF \\ $\sigma_s(t_k^-)$ \end{tabular} & \begin{tabular}[c]{@{}c@{}}Time \\difference \\ $t-t_k $\end{tabular} & $\sigma (t)$ \\ \hline
1, 2, 3 & -- & \begin{tabular}[c]{@{}c@{}}2 or 3, \\ 1 or 3, \\ 1 or 2\end{tabular} & -- & $>T_{D_1}$ & 1, 2, 3 \\ \hline
1, 2, 3 & -- & \begin{tabular}[c]{@{}c@{}}2 or 3, \\ 1 or 3, \\ 1 or 2\end{tabular} & 4 & $>T_{D_2}$ & 1, 2, 3 \\ \hline
1, 2, 3 & 4 & -- & -- & -- & 1, 2, 3 \\ \hline
4 & -- & -- & -- & -- & 4 \\ \hline
\end{tabular}
\end{table}

\subsection{MAPOF-based control}
By combining the MAPOFs \eqref{eq:forces} with the switching signal $\sigma:[0,\infty)\to\{1,2,3,4\}$ in \eqref{eq:switching_signal}, we obtain the final MAPOF-based control law
\begin{equation}
	\begin{split}
		u_{\sigma(t)}&=F_{\sigma(t)}(\xi)-k_d\dot{\xi},
	\end{split}
	\label{eq:control}
\end{equation}
where $F_{\sigma(t)}\in \mathbb{R}^2$ is the APOF in dependency of the switching signal $\sigma(t)$ and $k_d \in \mathbb{R}_{>0}$ is a positive gain, used to add dissipation in the velocity to the control. 

\subsection{Closed-loop system and error dynamics}
By using (\ref{eq:system_initial}), (\ref{eq:switching_signal}) and \eqref{eq:control}, we
obtain the closed-loop dynamics
\begin{equation}
	\begin{split}
		\ddot{\xi}&=u_{\sigma(t)},\\
		u_{\sigma(t)}&=F_{\sigma(t)}(\xi)-k_d\dot{\xi}.
	\end{split}
	\label{eq:system}
\end{equation}

To derive tuning conditions that ensure closed-loop stability, we consider the error dynamics of \eqref{eq:system} with respect to the target position. Recalling that $\eta\in \mathbb{R}^2$ is a constant point, the error states are given by
\begin{equation}
	\begin{split}
		x_1&=\eta-\xi \,\in \, \mathbb{R}^2 ,\\
		x_2&=-\dot{\xi}\, \in \,\mathbb{R}^2,
	\end{split}
	\label{eq:error}
\end{equation}
and the error state space model by
\begin{equation}
	\begin{split}
		&\dot{x}_1=x_2,\\
		&\dot{x}_2=-F_{\sigma(t)}(x_1)-k_dx_2.
	\end{split}	
	\label{eq:error_s}
\end{equation}
The error dynamics \eqref{eq:error_s} with switching signal $\sigma$ can be formulated compactly as the switched piecewise affine linear  system
\begin{equation}
	\dot{x}(t)=A_{\sigma(t)}x(t)+b_{\sigma(t)},
	\label{eq:system_AB}
\end{equation}
where $A_{\sigma(t)}$ and $b_{\sigma(t)}$ depend on the activation function (\ref{eq:switching_signal}) and the artificial forces (\ref{eq:forces}) in the following way:

\begin{equation}
\begin{split}
        A_1=\begin{bmatrix}
 0& 1 \\ 
 -k_{\eta}&-k_d 
\end{bmatrix} \otimes I_2, \,\, 
A_2=\begin{bmatrix}
 0& 1 \\ 
- k_g&-k_d 
\end{bmatrix}\otimes I_2, \\ 
A_3=\begin{bmatrix}
 0& 1 \\ 
- k_g&-k_d 
\end{bmatrix}\otimes I_2, \,\, 
A_4=\begin{bmatrix}
 0& 1 \\ 
k_{\zeta}&-k_{d}
\end{bmatrix} \otimes I_2,
\end{split} 
\label{eq:matrices}
\end{equation}

\begin{equation}
\begin{split}
       b_1 = \underline{0}_4 , \,\, 
b_2 = \begin{bmatrix}
\underline{0}_2\\-k_g\left(g_{1}^*-\eta\right) 
\end{bmatrix}, \\
b_3 = \begin{bmatrix}
\underline{0}_2\\-k_g\left(g_{2}^*-\eta\right) 
\end{bmatrix}, \,\,
b_4 = \begin{bmatrix}
\underline{0}_2\\k_{\zeta}\left(\zeta-\eta\right) 
\end{bmatrix},
\end{split}
\label{eq:b_matrices}
\end{equation} 
where $\otimes$ denotes the Kronecker product, $I_n$ denotes the $n \times n$ identity matrix, and $\underline{0}_n \in \mathbb{R}^n$ is the zero vector.

It can be easily verified that the matrices $A_i$, $i=1,2,3,$ are Hurwitz, while the matrix $A_4$ is unstable. The latter property is due to the repulsion force $F_4$ in \eqref{eq:forces}, which is crucially needed for the purpose of obstacle avoidance. Hence, in the present case, the unstable behavior in mode $4$ is indeed desired and imposed by design.

\section{DWELL-TIME DEPENDENT TUNING CONDITIONS FOR CLOSED-LOOP STABILITY}
This section is dedicated to the derivation of tuning conditions for the error dynamics \eqref{eq:error_s}, which guarantee closed-loop stability. This is a very challenging task for the following reasons. The dynamics \eqref{eq:system_AB}, \eqref{eq:switching_signal} is a switched, piecewise affine linear system. This implies that not all subsystems share the origin as a common equilibrium point, as is usually assumed \cite{allerhand2012robust}. In fact, the origin -- our desired equilibrium point -- is only an admissible equilibrium point for the dynamics $A_1,$ $b_1,$ see \eqref{eq:matrices}, \eqref{eq:b_matrices}. 

In addition, the dynamics \eqref{eq:system_AB} contain stable and unstable subsystems. 
To overcome these challenges, we employ tools from hybrid systems theory and impose a minimum dwell-time to limit the number of switches and, in particular, the time spent in the unstable mode.

\subsection{Existence of a unique equilibrium point}
The next lemma shows that the switching law \eqref{eq:switching_signal} ensures that the origin is the only admissible equilibrium of the switched system \eqref{eq:system_AB}, \eqref{eq:switching_signal}. Let $\zeta$ denote the obstacle position and $\eta$ the target position. Recall from Section~\ref{sec:prob} that $r_d$ is the UAV's detection radius.

\begin{lemma}
	Consider the switched system \eqref{eq:system_AB}, \eqref{eq:switching_signal}. Suppose that $\|\eta-\zeta\|>r_d.$ Then, the origin is the only admissible equilibrium point of \eqref{eq:system_AB}, \eqref{eq:switching_signal}.
	\label{lem:ex}
\end{lemma}
\begin{IEEEproof}
Any equilibrium point $x^\star$ of the system \eqref{eq:system_AB}, \eqref{eq:switching_signal} has to satisfy
\begin{equation*}
    \dot{x}^\star(t)=A_{\sigma(t)}x^\star(t)+b_{\sigma(t)} =0 \,\,\,\,\,\,  \forall t \geq t_0, \,\, \sigma_s\in\{1,2,3,4\}.
\end{equation*}
If we consider the individual subsystems in \eqref{eq:system_AB}, we easily deduce from the matrices \eqref{eq:matrices} and \eqref{eq:b_matrices} four possible equilibrium points, i.e.,
 \begin{equation}
     \begin{split}
         x^{1\star} &= \underline{0}_4, \\
         x^{2\star} &=\left[ \underline{0}_2^\top ,\left (\eta - g_1^* \right)\right]^\top ,\\
         x^{3\star} &=\left[ \underline{0}_2^\top ,\left (\eta - g_2^* \right)\right]^\top,\\
         x^{4\star} &=\left[ \underline{0}_2^\top ,\left (\eta - \zeta \right)\right]^\top.
     \end{split}
     \label{eq:equilibrium_p}
 \end{equation}
The points in \eqref{eq:equilibrium_p} in terms of the position of the UAV correspond to the target position, the virtual positions, and the obstacle. According to the activation function \eqref{eq:switching_signal}, we deduce that:
\begin{itemize}
    \item The points $x^{2\star}$ and $x^{3\star}$ belong to the subregion 1, cf. Fig.~\ref{fig:geo_diagram}. Hence, if initialized in either $x(t_0)=x^{2\star}$ or $x(t_0)=x^{3\star},$ according to \eqref{eq:switching_signal} after a time $t-t_0>T_{D_1}$, the system will switch to mode 1. Consequently, neither $x^{2\star}$ nor $x^{3\star}$ are equilibrium points of the system \eqref{eq:system_AB}, \eqref{eq:switching_signal}.
    \item The equilibrium point $x^{4\star}$ is not an {\em admissible} equilibrium point of the system dynamics, since this point corresponds to the obstacle position and the UAV cannot start in the obstacle.

    \item If $x(t_0)=x^{1\star}$, then by \eqref{eq:switching_signal} no further switches occur for all $t>t_0$, i.e., $x(t)=x^{1\star}$ for all $t\geq t_0$, which implies that $x^{1\star}$ is an equilibrium point of the system \eqref{eq:system_AB}, \eqref{eq:switching_signal}.

\end{itemize}  
\end{IEEEproof}

\subsection{Some preliminary observations}
Since the matrices $A_1$, $A_2,$ and $A_3$ in \eqref{eq:matrices} are given in companion form, it is immediate to see that they all have $2$ distinct eigenvalues. Hence, for any choice of gains $k_\eta,$ $k_g$ and $k_d$ there exist positive constants $a_i\in\R$ and $\alpha_i\in\R$, such that
\begin{equation}
\|e^{A_it}\|_2\leq e^{a_i-\alpha_it} \quad \forall t\geq0,
\end{equation}
where

\begin{equation}
    \begin{split}
        \alpha_1&= \min \left |  \textup{Re} \left\{ -\frac{k_d}{2} \pm \frac{\sqrt{k_d^2-4k_{\eta}}}{2}  \right\}   \right |, \\
        \alpha_2=\alpha_3 &= \min \left|  \textup{Re} \left\{ -\frac{k_d}{2} \pm \frac{\sqrt{k_d^2-4k_{g}}}{2}  \right\}  \right|.
    \end{split}
\end{equation}

Similarly, we can see that $A_4$ has always a positive and a negative eigenvalue. Hence, there exist positive constants $a_4\in\R$ and $\alpha_4\in\R$, such that
\begin{equation}
	\|e^{A_4t}\|_2\leq e^{a_4+\alpha_4t} \quad \forall t\geq0,
	\label{eq:A4}
\end{equation}

where,

\begin{equation}
    \begin{split}
        \alpha_4&= \max \left |  \textup{Re} \left\{ -\frac{k_d}{2} \pm \frac{\sqrt{k_d^2+4k_{\zeta}}}{2}  \right\} \right| \\
    \end{split}
\end{equation}
Let

\begin{equation}
\begin{split}
\alpha^-=\min_{1\leq i\leq 3} \alpha_i,\quad \alpha^+=\alpha_4, \quad a=\max_{1\leq i\leq 4} a_i.
\end{split} 
\end{equation} 
It follows from \eqref{eq:A4} together with \eqref{eq:switching_signal} that every time the unstable dynamics are activated, the activation time is finite. More precisely, suppose that $\sigma(t)=4$} and denote by $0<\mu <r_m$ the minimum distance of the UAV from the obstacle. 
Then, the time interval $T_t\in\R$, after which zone 4 is left (see Figure~\ref{fig:map_force}), can be computed from
\begin{equation}
		e^{a+\alpha^+T_t}\mu \leq r_m
\end{equation}
as		
\begin{equation}		
		0<T_t\leq\frac{\log\left(\frac{r_m}{\mu} \right)-a}{\alpha^+}.
		\label{eq:t_t}
\end{equation}
Clearly, in practice, $\mu \ll r_m.$

\subsection{Tuning criteria for MAPOFs control}
\label{sec:proof}

For the derivation of the tuning conditions, the tailored design of the switching function $\sigma$ in \eqref{eq:switching_signal} and the state space partitioning in Figure~\ref{fig:map_force} are exploited. These imply that only specific sequences of switches are permitted. In particular, whenever the unstable mode is active, a stable mode must be activated subsequently. 
 
Denote the minimum dwell-time for switching from a stable to a stable mode, i.e., $\sigma\in\{1,2,3\}$, by $T_{D_1}$ and the dwell-time after switching from the unstable mode, i.e., $\sigma=4$, to a stable mode by $T_{D_2}.$

We can now state the proposed tuning conditions.
\begin{theorem}
\label{th:main}
Consider the switched system \eqref{eq:system_AB}, \eqref{eq:switching_signal}. Its origin is a stable equilibrium point if the dwell-times $T_{D_1}$ and $T_{D_2}$ are chosen such that
\begin{equation}
    T_{D_1} \geq  \frac{a+\rho}{\alpha^-},
        \label{eq:td1}
\end{equation}
and
\begin{equation}
	T_{D_2} \geq \frac{2a+\alpha^+T_t+\rho}{\alpha^-},
	\label{eq:td2}
\end{equation}
where $\rho >0$ is a constant and $T_t > 0$ is defined in \eqref{eq:t_t}. 

\end{theorem}

\begin{IEEEproof} 
By invoking Lemma~\ref{lem:ex}, we conclude that the origin is the unique equilibrium point of the system \eqref{eq:system_AB}, \eqref{eq:switching_signal}.
Next, we rewrite the system \eqref{eq:system_AB}, \eqref{eq:switching_signal} in new coordinates as
\begin{align*}
	z=x-x_{0_p}, \,\,\,\,\, x_{0_p}=(-A_p)^{-1} b_p, \,\,\,\,\, p \in \{ 1,2,3,4\}.
\end{align*}
The new system generated is a hybrid system, i.e.,
\begin{equation}
	\begin{split}
		\dot{z}  &=A_{\sigma(t)}z , \,\,\,\, t_k\leq t \leq t_{k+1}, \\
		z(t_k)  &=z(t_k) + \delta_k, \,\,\,\,\delta_k=x_{0_{\sigma(t_k-1)}}-x_{0_{\sigma(t_k)}},
	\end{split}
\label{eq:hybrid_system}
\end{equation}
where $t_k \in \mathbb{R}$ is the point in time when the $k$-th switch occurs. By analyzing the behavior of the system at each instant of switching, an expression for the $k$-th switching is deduced as follows
\begin{equation}
	\begin{split}
		z(t_1)=&e^{A_{\sigma(t_0)}(t_1-t_0)}z(t_0)+\delta_1 ,\\
		z(t_2)=&e^{A_{\sigma(t_1)}(t_2-t_1)}z(t_1)+\delta_2, \\
		=&e^{A_{\sigma(t_1)}(t_2-t_1)}\left( e^{A_{\sigma(t_0)}(t_1-t_0)}z(t_0)+\delta_1\right)+\delta_2,\\
		z(t_3)=&e^{A_{\sigma(t_2)}(t_3-t_2)}e^{A_{\sigma(t_1)}(t_2-t_1)}e^{A_{\sigma(t_0)}(t_1-t_0)}z(t_0)\\&+e^{A_{\sigma(t_2)}(t_3-t_2)}e^{A_{\sigma(t_1)}(t_2-t_1)}\delta_1 \\
		&+e^{A_{\sigma(t_2)}(t_3-t_2)}\delta_2 +\delta_3,\\
		z(t_k)=&\left (\prod_{i=1}^{k}e^{A_{\sigma(t_{k-i})}(t_{k-i+1}-t_{k-i})} \right) z(t_0) \\&+ \sum_{i=1}^{k-1}\left (\prod_{j=i+1}^{k} e^{A_{\sigma(t_{k-j+i})}(t_{k-j+i+1}-t_{k-j+i}  )} \right) \delta_i \\
        & +\delta_k, \,\,\,\,\,\,\,\,k>1.    
	\end{split}
	\label{eq:z_tk}
\end{equation}

Recall from \eqref{eq:t_t} that by the design of the MAPOFs control \eqref{eq:control}, any activation of the unstable mode $\sigma=4$ is followed by the activation of a stable mode after no more than $T_t$ time units.

Hence, without loss of generality, it is assumed that at $t=t_k$ a stable mode is activated.

Next, we introduce the function $m(k,i) \in \mathbb{Z},\,\, \forall i<k$, that returns the number of activations of the unstable mode between the $i$-th and $k$-th switching. According to the above considerations, we can bound the second term of (\ref{eq:z_tk}) as
\begin{equation}
    \begin{split}
        \left \| \sum_{i=1}^{k-1}\left (\prod_{j=i+1}^{k} e^{A_{\sigma(t_{k-j+i})}(t_{k-j+i+1}-t_{k-j+i}  )} \right) \right \| \\
       \leq   \sum_{i=1}^{k-1}e^{\left(k-i-2m(k,i)\right)(a-\alpha^-T_{D_1})}e^{m(k,i)(a-\alpha^-T_{D_2})}\\e^{m(k,i)(a+\alpha^+T_{t})},
    \end{split}
    \label{eq:second_term_1}
\end{equation}
where the first exponential term corresponds to the possible switching between stable modes, the second exponential term to the switching from the unstable to a stable mode, and the third exponential term to the overall activation time of the unstable mode. By using the dwell-time restrictions defined in (\ref{eq:td1}) and (\ref{eq:td2}), the right-hand-side of the inequality \eqref{eq:second_term_1} is simplified to
\begin{equation}
    \sum_{i=1}^{k-1} e^{-\left(k-i-2m(k,i)\right)\rho} e^{-m(k,i)\rho}=\sum_{i=1}^{k-1} e^{-\left(k-i-m(k,i)\right)\rho}.
    \label{eq:second_term_2}
\end{equation}
Considering the nature of the switching of the MAPOFs control \eqref{eq:control}, there can be at most half of the total switches $k-i$ to the unstable mode.
Thus, we deduce that $m(k, i)\leq \frac{k-i}{2}$. By considering the above restriction in (\ref{eq:second_term_2}), we finally obtain
\begin{equation}
    \sum_{i=1}^{k-1} e^{-\left(k-i-m(k,i)\right)\rho} \leq \sum_{i=1}^{k-1} e^{-\left(\frac{k-i}{2}\right)\rho} = \frac{e^{-\left(\frac{k-1}{2}\right)\rho}-1}{1-e^{\frac{\rho}{2}}}.
    \label{eq:second_term_3}
\end{equation}

In a similar fashion, we obtain an upper bound for the first term of the expression for $z(t_k)$ in (\ref{eq:z_tk}):
\begin{equation}
\begin{split}
   &\left \|  \prod_{i=1}^{k}e^{A_{\sigma(t_{k-i})}(t_{k-i+1}-t_{k-i})}  \right \| \\ 
   &\leq e^{\left(k-2m(k,0)\right)(a-\alpha^-T_{D_1})}e^{m(k,0)(a-\alpha^-T_{D_2})}e^{m(k,0)(a+\alpha^+T_{t})} .
\end{split}
\label{eq:first_term}
\end{equation}
By considering the lower bounds for the dwell-times specified in (\ref{eq:td1}), (\ref{eq:td2}) and the fact that $m(k,0)\leq \frac{k}{2}$, we obtain
\begin{equation}
\begin{split}
   \left \|  \prod_{i=1}^{k}e^{A_{\sigma(t_{k-i})}(t_{k-i+1}-t_{k-i})}  \right \|  \leq e^{-\frac{k}{2}\rho}.
\end{split}
\label{eq:first_term_2}
\end{equation}

By combining (\ref{eq:second_term_3}) and (\ref{eq:first_term_2}) as well as denoting $\delta = \sup_{k\geq 0 } \delta_k $, see \eqref{eq:hybrid_system}, we obtain
\begin{equation}
\begin{split}
   \left \|  z(t_k)  \right \|  \leq e^{-\frac{k}{2}\rho}\left \|  z(t_0)  \right \| +\left( \frac{e^{-\left(\frac{k-1}{2}\right)\rho}-1}{1-e^{\frac{\rho}{2}}}\right)\delta +\delta.
\end{split}
\label{eq:final_bound}
\end{equation}
Consequently, the trajectories of the system (\ref{eq:hybrid_system}) are bounded and the origin is a stable equilibrium point of the system \eqref{eq:system_AB}, \eqref{eq:switching_signal}.

\end{IEEEproof}

\subsection{Dwell-time relaxation}
The dwell-time conditions in Theorem~\ref{th:main} may be restrictive, when the detection radius $r_d$ is small, since then they may delay the activation of the unstable mode $\sigma=4$ too much. Therefore, we seek to relax the dwell-time conditions established in Theorem~\ref{th:main} for that scenario. To this end, we consider the same setup as for Theorem~\ref{th:main}, but instead of setting a time restriction between the switching from a stable mode $\sigma=\{1,2,3\}$ to the unstable mode $\sigma=4$, we allow for an arbitrarily fast switching and compensate the time in the next stable mode. For example, in the present case, any of the stable modes $\sigma=\{1,2,3\}$ of the system \eqref{eq:system_AB}, \eqref{eq:switching_signal} can be used to compensate this fast switching. We extend the stability result in this regard.\\


\begin{corollary}
\label{cor:1}
    Consider the switched system \eqref{eq:system_AB}, \eqref{eq:switching_signal}. 
    For any switching between any two subsystems $\sigma \in \{1,2,3\}$, set the dwell-time $T_{D_1}$ as given in~\eqref{eq:td1}.  
    For the switching from any subsystem $\sigma \in \{1,2,3\}$ to the subsystem $\sigma=4,$ set $T_{D_1}=0.$ Furthermore, set $T_{D_2}$ as
    \begin{equation}
    T_{D_2} \geq \frac{3a+\alpha^+T_t+\rho}{\alpha^-},
   \label{eq:td3}
    \end{equation}
    where $0<\rho<a$ is a constant and $T_t > 0$ is defined in \eqref{eq:t_t}. Then, the origin is a stable equilibrium of the switched system~\eqref{eq:system_AB}, \eqref{eq:switching_signal}.\\
\end{corollary}

\begin{IEEEproof}
Considering the switching logic in~\eqref{eq:switching_signal}, any activation of the unstable mode, i.e., $\sigma=4$, is preceded and followed by an activation of a stable mode, i.e., $\sigma\in\{1,2,3\}.$ 
Hence, for any time interval $[t_i,t_k]\subset\R,$ $i<k$, we introduce two functions $v(k,i)$ and $q(k,i)$, where $v(k,i)$ denotes the number of activations of stable modes with dwell-time $T_{D_1}$, see \eqref{eq:td1} and $q(k,i)$ is the number of sequences of stable-unstable-stable modes during the considered time interval. Moreover, we restrict the analysis to time instants $t_k$, which satisfy $k=v(k,0)+3q(k,0)$. This implies that we only consider such time instants $t_k$, where any sequence of stable-unstable-stable modes has been completed. As argued before, see \eqref{eq:t_t}, any activation of the unstable mode $\sigma=4$ is only of a finite-time duration. Hence, the system \eqref{eq:system_AB}, \eqref{eq:switching_signal} will always exhibit a $t_k$ satisfying $k=v(k,0)+3q(k,0)$.

Now, we can bound the first term of $z(t_k)$ in \eqref{eq:z_tk} as
\begin{equation}
\begin{split}
   &\left \|  \prod_{i=1}^{k}e^{A_{\sigma(t_{k-i})}(t_{k-i+1}-t_{k-i})}  \right \|\leq e^{v(k,0)(a-\alpha^-T_{D_1})}\cdot\\&\left(e^{q(k,0)(a-\alpha^-T_{D_1})}e^{q(k,0)(a+\alpha^+T_{t})}e^{q(k,0)\left(a-\alpha^-T_{D_2}\right)}\right)\\
   &\leq e^{-v(k,0)\rho}e^{-q(k,0)\rho}=e^{-\left(v(k,0)+q(k,0)\right)\rho},
\end{split}
\label{eq:first_term_coro_2}
\end{equation}
where we have used $T_{D_1}$ in \eqref{eq:td1} and $T_{D_2}$ in~\eqref{eq:td3} together with the fact that, by the conditions of the corollary, $T_{D_1}=0$ for terms having $q(k,0)(a-\alpha^-T_{D_1})$ in the exponent to obtain the last inequality.

For the second term in \eqref{eq:z_tk}, since the grouping of stable-unstable-stable switching sequences needs at least a group of three switchings, by recalling $\delta = \sup_{k\geq 0 } \delta_k $, see \eqref{eq:hybrid_system}, we rewrite this term as
\begin{equation}
    \begin{split}
&\sum_{i=1}^{k-1}\left (\prod_{j=i+1}^{k} e^{A_{\sigma(t_{k-j+i})}(t_{k-j+i+1}-t_{k-j+i}  )} \right) \delta_i+\delta_k \\
&=\sum_{i=1}^{k-3}\left (\prod_{j=i+1}^{k} e^{A_{\sigma(t_{k-j+i})}(t_{k-j+i+1}-t_{k-j+i}  )} \right) \delta_i\\
&+e^{A_{\sigma(t_{k-1})}(t_k-t_{k-1})}e^{A_{\sigma(t_{k-2})}(t_{k-1}-t_{k-2})}\delta_{k-2}\\
		&+e^{A_{\sigma(t_{k-1})}(t_k-t_{k-1})}\delta_{k-1} +\delta_k\\
  &\leq \sum_{i=1}^{k-3}e^{v(k,i)(a-\alpha^-T_{D_1})}e^{q(k,i)(a-\alpha^-T_{D_1})(a+\alpha^-T_{t})(a-\alpha^-T_{D_2})} \delta \\
&+e^{-\rho}\delta+e^{-\rho}\delta +\delta\\
    &\leq \left( \sum_{i=1}^{k-3}e^{-(v(k,i)+q(k,i))\rho}+2e^{-\rho}+ 1\right)\delta,
    \end{split}
    \label{eq:second_term_cor_1}
\end{equation}
where we have used again $T_{D_1}$ in \eqref{eq:td1} and $T_{D_2}$ in~\eqref{eq:td3} together with the conditions of the corollary as well as the fact that we only consider time instants $t_k$, which satisfy $k=v(k,0)+3q(k,0).$

Since for any $k$ and $i$, $v(k,i)+q(k,i)\leq \frac{k-i}{3},$ we have that
\begin{equation}
    \begin{split}
\sum_{i=1}^{k-3}e^{-(v(k,i)+q(k,i))\rho}&\leq \sum_{i=1}^{k-3}e^{-\left(\frac{k-i}{3}\right)\rho}\\
&=\frac{e^{-\frac{\rho}{3}(k-1)}-e^{-\frac{2}{3}\rho}}{1-e^{\frac{\rho}{3}}}.
    \end{split}
    \label{eq:second_term_cor_2}
\end{equation}
Hence, by introducing the constant

\begin{equation}
    c=e^{-2\rho}+e^{-\rho}
\notag 
\end{equation}
and combining \eqref{eq:first_term_coro_2}, \eqref{eq:second_term_cor_1} with \eqref{eq:second_term_cor_2}, we obtain
\begin{equation}
\begin{split}
   \left \|  z(t_{k})  \right \|  \leq e^{-\frac{k}{3}\rho}\left \|  z(t_0)  \right \| + \left(\frac{e^{-\frac{\rho}{3}(k-1)}-e^{-\frac{2}{3}\rho}}{1-e^{\frac{\rho}{3}}}+c+1\right)\delta.
\end{split}
\label{eq:final_bound_coro2}
\end{equation}

 Therefore, the trajectories of the system~\eqref{eq:hybrid_system} are bounded, and the origin of the system \eqref{eq:system_AB}, \eqref{eq:switching_signal} is a stable equilibrium point.

\end{IEEEproof}

\begin{figure}
     \centering
     \includegraphics[scale=0.6]{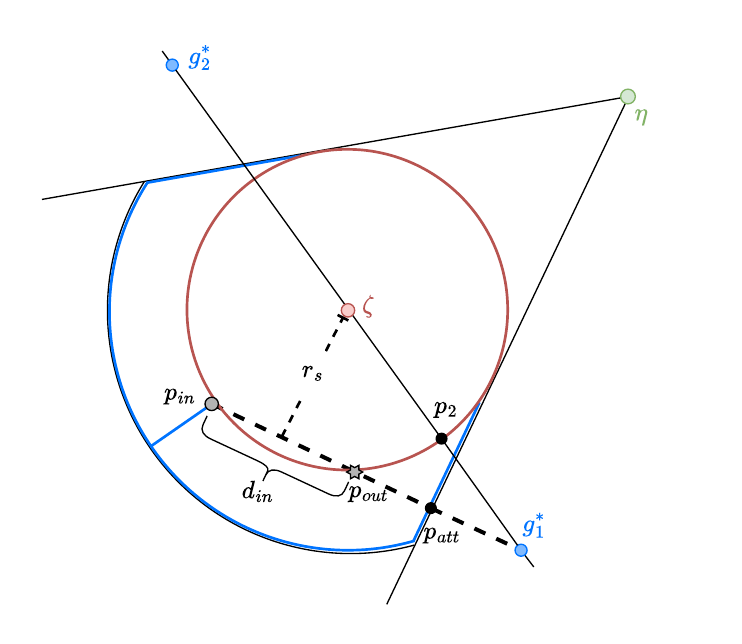}
     \caption{Illustration of the scenario considered in Corollary~\ref{cor:1}. The switching after the unstable mode $\sigma=4$ is restricted by the dwell-time $T_{D_2}$ in \eqref{eq:td3}, so that the UAV maintains the avoidance force $\sigma=2$ until the point $g_{1}^{*}$.}
     \label{fig:diagram_in}
 \end{figure}

 A consequence of holding the dwell-time restriction after the unstable mode is that, in the worst case, the UAV enters the repulsion zone again without the repulsion mode $\sigma=4$ being immediately activated. However, this is not a dangerous maneuver, since either of the avoidance modes $\sigma=2$ or $\sigma=3$ drive it out of the repulsion zone in finite time.
 This scenario is illustrated in Figure~\ref{fig:diagram_in}. In this case, the UAV is situated at the intersection of three zones denoted as $p_{in} \in \mathbb{R}^2$ in Figure~\ref{fig:diagram_in}. When the UAV applies the avoidance force, it needs to cover a defined distance denoted as $d_{in}>0$. The position of the UAV when it is outside of the repulsion zone is denoted as $p_{out}\in \mathbb{R}^2$, and the radius that guarantees avoidance with the obstacle is $r_{s}>0$. According to the MAPOFs design, it is deduced the limits $d_{in}<\sqrt{2}r_m$ and $\frac{r_m}{\sqrt{2}}<r_{s}<r_m$. Then, the time that the UAV takes to cross the repulsion $T_{in}$ can be computed from 
\begin{equation}
    \left \| p_{in}-g_{1}^* \right \| e^{a-\alpha^-T_{in}}\geq \left \| p_{in}-g_{1}^* \right \| -d_{in}
\end{equation}
as
\begin{equation}
    T_{in} \leq \frac{a-\log \left(1-\frac{d_{in}}{ \left \| p_{in}-g_{1}^* \right \|} \right) }{\alpha^-}.
   \label{eq:tin}
\end{equation}

In practice, $d_{in} < \left \| p_{in}-g_{1}^* \right \|$, and $T_{D_2}$ is designed as $T_{D_2}>T_{in}$. Indeed, the UAV holds the avoidance mode also if it is in the repulsion zone and it maintains at least a security area with respect to the obstacle  $r_s$. \\

    \textbf{Remark 1.} The dwell-time $T_{D_2}$ provides a relaxation of the initial dwell-time definition for taking advantage of the proposed switched system~\eqref{eq:system_AB}, \eqref{eq:switching_signal} and improving collision avoidance. The UAV starts to avoid obstacles, when it enters zones 2 or 3. If the avoidance zone is insufficient, the UAV enters the repulsion zone 4, where the force is opposite to the obstacle position. Eventually, the UAV switches back to subsystem 2 or 3, where it compensates for the time of the last switches. 
    In principle, the UAV could already switch at the point $p_{att}$ to the subsystem 1, since by the definition of the MAPOF, after crossing the point $p_{att}$, the UAV will not enter again the repulsion zone.  \\

\subsection{Multiple obstacles analysis}

The last proof focuses on a $k$-th switch and only considers that the dwell-time restrictions or dwell-time relaxations hold. Furthermore, the time restrictions only depend on the matrices $A_p$. The obstacle's position, the temporal restrictions, and the goal are related to $\delta$. So, the stability analysis remains valid for an arbitrary number of obstacles if always the same switching strategy~\eqref{eq:switching_signal} is applied. The remaining problem is the intersection of the different zones, when the obstacles are close. One remedy to this scenario is to restrict the minimal distance between obstacles to $r_m+r_d$. Alternatively, if multiple obstacles are close to each other, we propose to calculate the center position and use this point as a unique obstacle position to generate the MAPOFs map. 

\section{NUMERICAL CASE STUDIES}

Two case studies are explored to validate the algorithm. The first scenario involves a single obstacle, while the second scenario considers multiple obstacles and different initial conditions. Every obstacle is modeled for both scenarios using a single point with a security radius $r_m = 3 \,\mathrm{m}$ that corresponds to the area of the unstable subsystem, as shown in Figure \ref{fig:map_force}. The UAV has a detection radius $r_d = 8\,\mathrm{m}$, and the virtual positions $g_{1}^{*}$ and $g_{2}^{*}$ are at $4.5 \, \mathrm{m}$ of the obstacles. 

The dwell-time values $T_{D_1}$ and $T_{D_2}$ are dependent on the eigenvalues of the matrices $A_p, \,\,p \in 1,2,3,4$, which in turn depends on the tuning parameters. Those parameters have been selected to reduce the dwell-time while ensuring that the attraction gains do not interfere with the avoidance. The parameters chosen for this purpose are  $k_\eta = 3.77$, $k_\zeta = 20$, $k_d = 4.37$, and $k_g = 1.90$. Assuming that the minimum distance between the UAV and the obstacle is $\mu = 5$ cm, the time for the UAV to go outside of the repulsion area is  $T_t = 0.5 \,\mathrm{s}$. In addition, selecting  $\rho = 0.1$ the dwell-times are $T_{D_1} = 1.60 \,\mathrm{s}$, $T_{D_2} = 6.25 \,\mathrm{s}$.

 \begin{figure}[t]
     \centering
     \includegraphics[scale=0.6]{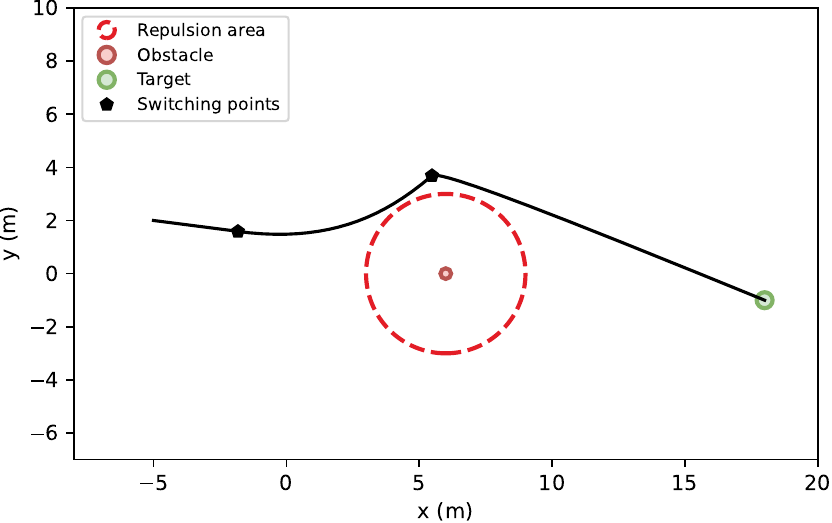}
     \caption{UAV trajectory with MAPOF strategy for collision avoidance in scenario 1: one obstacle.}
     \label{fig:scenario1}
 \end{figure}
  \begin{figure}[t]
     \centering
     \includegraphics[scale=0.6]{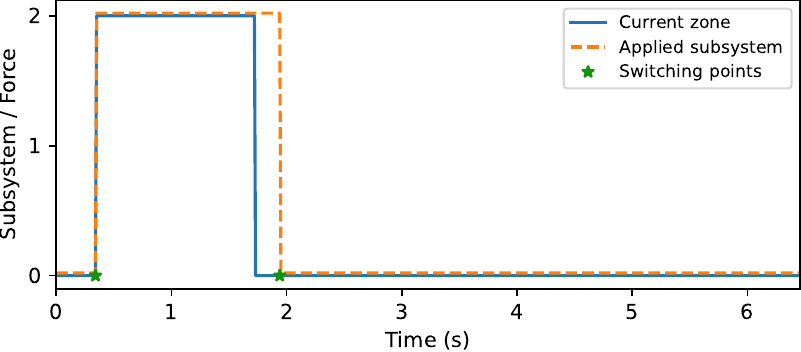}
     \caption{Switches in the different subsystems over time for scenario 1.}
     \label{fig:scenario1_switched}
 \end{figure}
\subsection{Scenario 1}
The first scenario considers an obstacle in $\zeta_1 = [6,0]^\top$, a target position in $\eta = [18,-1]^\top$ and, an initial condition in $\xi_0 = [-5,2]^\top$. Figure \ref{fig:scenario1} shows the behavior of the UAV under the MAPOF switching law in \eqref{eq:system_AB}, \eqref{eq:switching_signal}. The UAV avoids obstacles and reaches the target position. The pentagon shows the place where the switch occurs, and Figure \ref{fig:scenario1_switched} the subsystem active in each time instant. The UAV starts outside of the detection range of the obstacle, which means that the applied mode is the attraction to the obstacle. After 0.33 seconds, the UAV detects the obstacle and switches to avoidance mode. At at time of 1.71 seconds, the UAV changes to the attraction zone again. However, the dwell-time restriction maintains avoidance. Only at time 1.93 seconds, when the dwell-time restriction $T_{D_1}$ is achieved, the UAV switches to the attraction mode to reach the goal position. In this case, the UAV switches between stable modes, so the first dwell-time restriction is applied. Additionally, the test shows a smooth trajectory, avoiding the obstacle and the chattering effect. 

  \begin{figure}[t]
     \centering
     \includegraphics[scale=0.6]{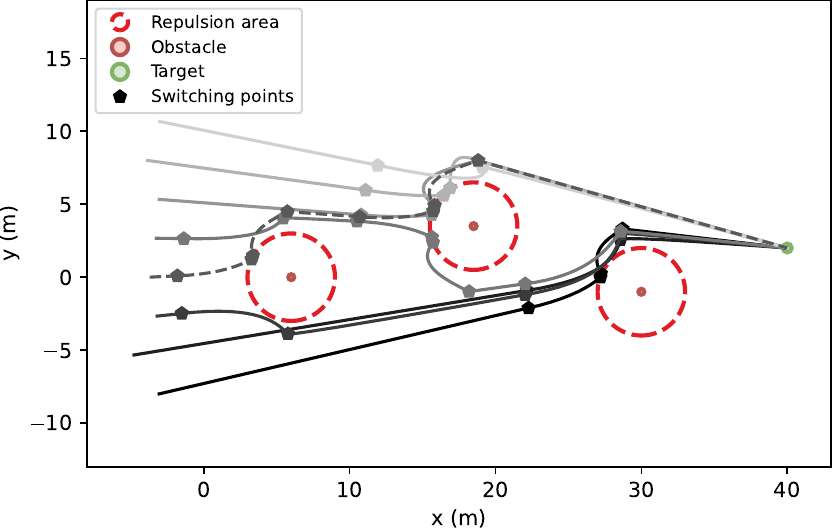}
     \caption{UAV trajectories using MAPOFs strategy for collision avoidance in scenario 2: Multiple obstacles and multiple initial conditions.}
     \label{fig:scenario2}
 \end{figure}

   \begin{figure}[t]
     \centering
     \includegraphics[scale=0.6]{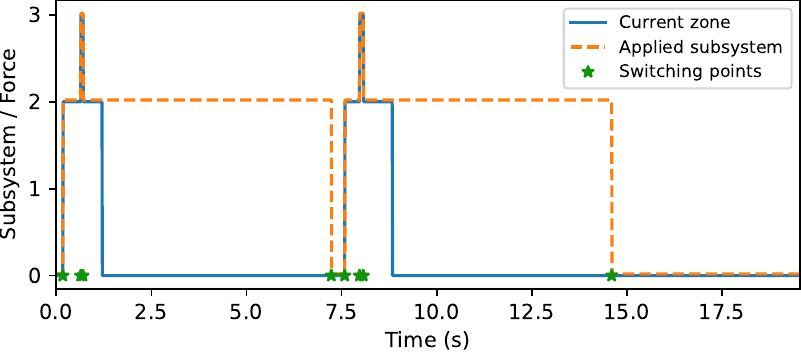}
     \caption{Switches in the different subsystems over time for scenario 2 for the dotted trajectory of Figure \ref{fig:scenario2}.}
     \label{fig:scenario2_switched}
 \end{figure}

 \subsection{Scenario 2}

In the second scenario, two new obstacles are added to the first scenario at positions $\zeta_2 = [18.5,3.5]^\top$ and $\zeta_3 = [30.0,-1.0]^\top$. The target position is $\eta = [40,2]^\top$, and eight initial positions are selected on the left side of the obstacles. Figure \ref{fig:scenario2} illustrates the UAV's behavior in this scenario. The UAV successfully avoids obstacles in all initial conditions while reaching the target position.

To analyze the behavior of the MAPOFs \eqref{eq:system_AB}, \eqref{eq:switching_signal} in the presence of multiple obstacles, we focus on the initial condition represented by the dotted line in Figure \ref{fig:scenario2}. Figure \ref{fig:scenario2_switched} presents the subsystem active in each time instant. The UAV switches eight times, sometimes with high frequency, indicating that the restriction of dwell-time $T_{D_1}$ does not hold. To compensate, the UAV remains in subsystem two until the dwell-time $T_{D_2}$ is satisfied, as shown in Figure \ref{fig:scenario2_switched}. In this case, the relaxation time is applied in the presence of unstable subsystems. For the other initial conditions, the relaxation permits the improvement of the smoothness of the UAV trajectories and collision avoidance while ensuring stability around the equilibrium point of the closed-loop system.

\section{Conclusion}
In this paper, a novel method for obstacle avoidance in UAV systems is proposed. The method guarantees a smooth trajectory and prevents chattering and deadlock issues. Our obstacle avoidance strategy involves an innovative approach to artificial potential functions, where the system switches between attraction and repulsion forces. By using a proposed switched law that involves concepts of dwell-time restrictions, we ensure the stability of the equilibrium point of the closed-loop system. We outline the tuning conditions required to maintain stability at the equilibrium point.

Additionally, we relax the conditions in scenarios where the constraints of the switched law are challenging to uphold while ensuring stability. We also present simulations to demonstrate the strategy's performance. The results show that in the presence of one or multiple obstacles, the UAV avoids the obstacles with a smooth trajectory.

Future work involves deriving tuning conditions for the asymptotic stability of the equilibrium point using the MAPOF strategy. Besides, for UAV implementation, it is important to consider errors and noise in measuring the position of the UAV or obstacles, which implies that the analysis of robustness is critical.


\bibliographystyle{IEEEtran}
\bibliography{bibliography}
\nocite{*}


 





\end{document}